\documentstyle[aps]{revtex}
\input epsf
\newcommand{\be}{\begin{equation}}
\newcommand{\ee}{\end{equation}}
\newcommand{\bea}{\begin{eqnarray}}
\newcommand{\eea}{\end{eqnarray}}

\begin{document}

\title{Scaling regimes for second layer nucleation}
\author{Joachim Krug}
\address{Fachbereich Physik, Universit\"at GH Essen,
D-45117 Essen, Germany }
  \date{\today}

\maketitle

\begin{abstract}
Nucleation on top of two-dimensional islands 
with step edge barriers is investigated
using scaling arguments. The nucleation rate is expressed in
terms of three basic time scales: The time interval between
deposition events, the residence time of atoms on the island,
and the encounter time required for $i^\ast + 1$ atoms forming
a stable nucleus to meet. Application to the problem of second-layer
nucleation on growing first layer islands yields a sequence of
scaling regimes with different scaling exponents relating
the critical island size, at which nucleation takes place, to the
diffusion and deposition rates. 
Second layer nucleation is fluctuation-dominated, in the sense
that the typical number of atoms on the island is small compared
to $i^\ast + 1$, when the first layer island density exponent 
$\chi$ satisfies $\chi (i^\ast + 1) < 2$.
The upper critical nucleus size, 
above which the conventional
mean-field theory of second layer nucleation is valid,
increases with decreasing dimensionality.
In the related case
of nucleation on top of multilayer mounds fluctuation-dominated
and mean-field like regimes coexist for arbitrary values of the
critical nucleus size $i^\ast$.


\end{abstract}

\section{Introduction}

Atomistic nucleation theory was developed in the sixties and seventies
in response to the realization
that critical nuclei of atomic dimensions
are common in far from equilibrium thin film growth 
\cite{walton62,zinsmeister,venables}. During the past decade the
subject has experienced a vigorous renaissance, driven by
the invention of atomic scale imaging techniques as well as
by the availability of realistic large scale simulations
\cite{brune98}. While this has resulted in a wide range of new
theoretical developments, the classical theory has stood the test
of time remarkably well. This is surprising because classical nucleation
theory relies, in the jargon of statistical mechanics, on an
approximation of {\it mean field} type, in the sense that it considers only
the evolution of spatially averaged concentrations of the different
surface species -- adatoms and clusters -- and thereby neglects effects
of spatial correlations and fluctuations.

It was recently pointed out that the classical
theory manifestly fails in describing nucleation events on top
of islands bounded by strong step edge barriers \cite{rottler99,krug00}. 
The onset of {\it second layer nucleation} on top of the first layer
islands is an important factor determining the growth mode and morphology
of multilayer films. Tersoff and coworkers \cite{tersoff94} developed
a theory of second layer nucleation based on the classical mean
field estimate
\begin{equation}
\label{mfnuc}
\omega_{\rm mf} \sim \nu A n^{i^\ast + 1}
\end{equation}
for the rate $\omega$ of nucleation events on the island,
in terms of the in-layer hopping rate
$\nu$, the island area $A$,
the adatom density $n$ and the size $i^\ast$ of the critical
(= largest unstable) cluster. In Refs.\cite{rottler99,krug00} it was
shown that for $i^\ast = 1$, 2 the mean field estimate disagrees with
microscopic considerations \cite{villain,politi96,politi97} 
and computer simulations.
The failure of mean field theory was qualitatively
attributed to the small number of atoms typically present on the island, and the associated large fluctuations in the occupancy of the island.

The question then arises whether mean field behavior may be recovered
at larger values of the critical nucleus size, where nucleation
necessarily involves a greater number of atoms, and thus fluctuation
effects should be reduced. A partial answer to this question was provided
in recent work by Heinrichs, Rottler and Maass (HRM) \cite{heinrichs00}.
They distinguish between two different nucleation modes:
In the {\it fluctuation-dominated} mode the mean number of adatoms
on the island (in the quasi-stationary state prior to nucleation) is much
smaller than the number $i^\ast + 1$ required for nucleation, and 
therefore nucleation is a rare, large fluctuation, while in the 
{\it mean field} regime, which corresponds to the classical approach of 
\cite{tersoff94}, the mean number of adatoms exceeds $i^\ast + 1$.
Consistency considerations then show that second layer nucleation
on top of first layer islands is fluctuation-dominated for $i^\ast \leq 2$
and mean-field-like for $i^\ast \geq 3$.

In this paper I elaborate on the observations of HRM, with the 
goal of further clarifying the role of fluctuations in second
layer nucleation. I begin by rederiving the main results of HRM
within the approach of \cite{krug00}, which relies on expressing
the nucleation rate in terms of the basic time scales of the process. While
less quantitative than the treatment of \cite{heinrichs00},
this provides, perhaps, a somewhat more transparent
and unified view. A simple, universal criterion for fluctuation-dominated
second layer nucleation is found which involves the first layer
island density exponent and the critical nucleus size (Eq.(\ref{genbound})).
I then extend the analysis to $d$-dimensional nucleation, finding the
same general structure as in the two-dimensional case, but
with a decreasing influence of fluctuations with increasing
$d$; as a byproduct,
I recover the expression for the 
one-dimensional first layer island density
exponent recently derived in Ref.\cite{kallabis98}. Finally, 
I generalize the theory of nucleation on top of multilayer mounds    
developed in \cite{krug00} to $i^\ast > 1$. In contrast
to the case of nucleation on first layer islands, here fluctuation-dominated
and mean field like nucleation regimes are found to coexist for arbitrary
values of $i^\ast$. This illustrates that the importance
of fluctuations depends crucially on the growth situation.

\section{Stationary rates of fluctuation-dominated nucleation}
\label{Stationary}

We consider a two-dimensional island of fixed linear extension $L$.
Atoms are deposited onto the island at rate $F$, they migrate on the
island with an in-layer hopping rate $\nu$ and descend from the island
at the inter-layer hopping rate $\nu'$. The mean time interval between
subsequent arrivals of atoms on the island is then
\begin{equation}
\label{deltat}
\Delta t \sim \frac{1}{F L^2}
\end{equation}
and the mean residence time of an atom on the island
in steady state 
is of the form \cite{rottler99,krug00,heinrichs00}
\begin{equation}
\label{tau}
\tau = \frac{a L^2}{\nu} + \frac{b L}{\nu'},
\end{equation}
where $a$ and $b$ are geometry-dependent constants. The first
term on the right hand side of (\ref{tau}) is of the order  
of the traversal time $\tau_{\rm tr} \sim L^2/\nu$ required
for the atom to migrate across the island, while the second term describes
the increase in the residence time due to the step edge barriers.
The second term dominates if $L \ll l_{\rm ES}$, where
\begin{equation}
\label{lES}
l_{\rm ES} \approx \nu/\nu'
\end{equation}
is the Ehrlich-Schwoebel length \cite{politi96}.

The stationary 
mean number of atoms on the island is given by $\tau/\Delta t$
\cite{krug00,heinrichs00},
and hence the condition
for the fluctuation-dominated mode of HRM reads
\begin{equation}
\label{fluct}
\frac{\tau}{\Delta t} \ll 1.
\end{equation}
It will turn out to be useful to write
this in the form $L \ll L_F$, where the flux scale 
$L_F$ is given by 
\begin{equation}
\label{LF}
L_F \sim \left\{ \begin{array}{l@{\quad:\quad}l}
(\nu'/F)^{1/3} & L \ll l_{\rm ES} \\ 
(\nu/F)^{1/4}  & L \gg l_{\rm ES}. \end{array} \right.
\end{equation}
For given $L$, $\nu$ and $\nu'$, Eq.(\ref{fluct}) can always be
satisfied by making the flux sufficiently small.

Nucleation with a critical nucleus size $i^\ast$ is treated within 
the ``noninteracting particle model'' \cite{heinrichs00}, in which
the lifetimes of all unstable clusters 
are assumed comparable to the inverse hopping rate $1/\nu$. 
Then the third relevant time scale in the problem, in addition to 
(\ref{deltat}) and (\ref{tau}), is given by
the {\it encounter time} $\tau_{\rm enc}$ required for 
the $i^\ast + 1$ atoms to find each other on the island. 
In order of magnitude we
have \cite{heinrichs00}
\begin{equation}
\label{tauenc}
\tau_{\rm enc} \sim L^{2i^\ast}/\nu.
\end{equation}
A simple derivation is given below in Section \ref{Dimension}.

Under the condition (\ref{fluct}), each nucleation event is uniquely
associated with the deposition of the $i^\ast + 1$'th atom. 
The number of nucleation events per unit time can then be written
as 
\begin{equation}
\label{omegadef}
\omega = p_{\rm nuc}/\Delta t, 
\end{equation}
where $p_{\rm nuc}$ 
is the probability for a 
freshly deposited atom to participate in a nucleation event.
The latter is generally of the form 
\begin{equation}
\label{pnuc}
p_{\rm nuc} = p_{i^\ast + 1} \cdot  p_{\rm enc},
\end{equation} 
where $p_{i^\ast + 1}$ denotes the probability that $i^\ast$ atoms were
present on the island prior to the arrival of the $i^\ast + 1$'th,
and $p_{\rm enc}$ is the probability that the $i^\ast + 1$ atoms encounter
each other before one of them leaves the island again. 
The quantity $p_{i^\ast + 1}$ is given by \cite{krug00}
\begin{equation}
\label{piast}
p_{i^\ast + 1} = \frac{1}{i^\ast !} \left( \frac{\tau}{\Delta t} \right)^{i^\ast}.
\end{equation}
The encounter probability can be estimated as \cite{heinrichs00}
\begin{equation}
\label{penc}
p_{\rm enc} \approx 1 - e^{-\tau/\tau_{\rm enc}}
\end{equation}
where for simplicity 
we have equated the residence time of the $i^\ast + 1$ atoms with 
that of a single atom (in fact the two differ by a factor $i^\ast + 1$). 
Thus $p_{\rm enc}$ is of order unity for $\tau/\tau_{\rm enc} \gg 1$
and of the order $p_{\rm enc}\approx \tau/\tau_{\rm enc}$ when
$\tau/\tau_{\rm enc} \ll 1$.

In the case $i^\ast = 1$ the encounter time (\ref{tauenc}) 
is of the same order
as the traversal time $L^2/\nu$, whereas the 
residence time (\ref{tau})
is of the order of  $\tau_{\rm tr}$ or larger. Thus one always has
$p_{\rm enc} = {\cal O}(1)$ and $p_{\rm nuc} \approx p_2$ \cite{krug00}. 
In contrast, for $i^\ast > 1$ it is possible to have a transition from
$\tau \gg \tau_{\rm enc}$ to $\tau \ll \tau_{\rm enc}$ with increasing island
size. This obviously requires that $\nu/\nu' \gg 1$, so that $\tau \sim L/\nu' \gg
\tau_{\rm enc}$ for small islands, and occurs at the characteristic length scale
\begin{equation}
\label{Lx}
L_\times \sim (\nu/\nu')^{1/(2 i^\ast -1)}.
\end{equation}
We note that this is small compared to the Ehrlich-Schwoebel length, 
and therefore the transition to 
{\it encounter-limited} nucleation occurs before the step edge
barriers become irrelevant, in the sense that $\tau$ becomes of the same
order as $\tau_{\rm tr}$. 

In the encounter-limited regime the 
expression (\ref{omegadef}) agrees in order of magnitude with the
mean field estimate (\ref{mfnuc}). To see this, note that
the adatom density on the island is of order $\tau/(\Delta t L^2) =
F \tau$,
and hence
\begin{equation}
\label{mfnuc2}
\omega_{\rm mf} \sim \nu L^2 \left( \frac{\tau}{L^2 \Delta t}
\right)^{i^\ast + 1} \approx \frac{1}{\Delta t}
\left( \frac{\tau}{\Delta t}
\right)^{i^\ast} \frac{\tau}{\tau_{\rm enc}}
\approx \frac{p_{i^\ast + 1} \cdot  p_{\rm enc}}{\Delta t}.
\end{equation}
In this sense the mean field approach remains applicable also outside
of its strict range of validity \cite{heinrichs00}.

We summarize the preceding considerations  in 
the following three expressions for the 
fluctuation-dominated nucleation 
rate, which apply with increasing
island size $L$:
\begin{equation}
\label{omega1}
\omega \approx \frac{1}{\Delta t} \left( \frac{\tau}{\Delta t} \right)^{i^\ast} \sim
F (F/\nu')^{i^\ast} L^{3 i^\ast + 2} \;\;\;\;\;\;\; L \ll L_\times \;\;
({\rm regime \; II})
\end{equation}
\begin{equation}
\label{omega2}
\omega \approx \frac{1}{\tau_{\rm enc}} \left( \frac{\tau}{\Delta t} 
\right)^{i^\ast+1}
\sim
F (F/\nu')^{i^\ast} (\nu/\nu')  L^{i^\ast + 3} \;\;\;\;\;\;\; L_\times \ll L
\ll l_{\rm ES} \;\; ({\rm regime \; III})
\end{equation}
\begin{equation}
\label{omega3}
\omega  \approx \frac{1}{\tau_{\rm enc}} \left( \frac{\tau}{\Delta t} 
\right)^{i^\ast+1}
\sim
F (F/\nu)^{i^\ast} L^{2 i^\ast + 4} \;\;\;\;\;\;\; L
\gg l_{\rm ES} \;\; ({\rm regime \; IV}).
\end{equation}
Regime III does not exist for $i^\ast = 1$. For $i^\ast > 1$, 
the encounter-limited regimes
III and IV extend also into the mean field realm, while 
regime II occurs only in the fluctuation-dominated nucleation mode.

Note that Eqs.(\ref{omega1}) and
(\ref{omega2}) correspond to the regimes II and I identified in \cite{krug00}
for $i^\ast = 2$. Here we prefer to enumerate the regimes in accordance with 
HRM \cite{heinrichs00}. Their regime I will be dealt with below in Section 
\ref{Nonstat}.

\section{Consistent scaling regimes in two dimensions}
\label{Consistent}

While Eqs.(\ref{omega1},\ref{omega2},\ref{omega3}) provide, in the specified
ranges of parameters, valid expressions for the nucleation rate on top of an island of given size, in their application to a specific growth situation additional consistency requirements arise \cite{heinrichs00}: 
When a given expression for $\omega$ is used to 
compute the island size at which nucleation occurs, it has to be checked that
this island size does indeed lie in the appropriate range. 

In this section we derive the resulting scaling regimes for second layer
nucleation in two dimensions. Following
earlier work \cite{rottler99,krug00,tersoff94,heinrichs00} we assume
a population of equal sized first layer
islands of density $N$. The island size
increases with time $t$ or coverage
$\theta = Ft$ according to the growth law
\begin{equation}
\label{growth}
L(t) \sim \sqrt{\theta/N}.
\end{equation}
For a nucleation rate of the form 
\begin{equation}
\label{Omega}
\omega = F \Omega L^k 
\end{equation}
the critical island size at which a second layer nucleus forms is then 
given by \cite{krug00}
\begin{equation}
\label{Lc}
L_c \sim (N \Omega)^{-1/(k+2)}.
\end{equation}
We further introduce the notation \cite{heinrichs00}
\begin{equation}
\label{gammaalpha}
\Gamma \equiv \nu/F, \;\;\;\;\; \alpha \equiv 
\nu'/\nu = l_{\rm ES}^{-1},
\end{equation}
where usually $\Gamma \gg 1$, $\alpha \ll 1$.
The first layer island density is then of the order of
\begin{equation}
\label{N}
N \sim \Gamma^{-\chi},
\end{equation}
which defines the island density exponent $\chi$.
In standard nucleation theory \cite{venables}
it is given by the expression
\begin{equation}
\label{chiconv}
\chi = \frac{i^\ast}{i^\ast + 2}.
\end{equation}
At the moment we will however leave $\chi$ 
unspecified to allow for the
possibility of different first layer nucleation mechanisms.
Generalizations of (\ref{chiconv}) will be derived in Section
\ref{Dimension}.

\subsection{Stationary nucleation regimes}
\label{Stat}

Inserting Eqs.(\ref{omega1},\ref{omega2},\ref{omega3}) into (\ref{Lc}) one
obtains expressions for the critical island size of the general form 
\cite{heinrichs00}
\begin{equation}
\label{Lcgen}
L_c \sim \Gamma^{\gamma} \alpha^{\mu} ,
\end{equation}
where the exponents $\mu$ and $\gamma$ are given by 
\begin{equation}
\label{exp1}
\gamma = \frac{\chi + i^\ast}{3 i^\ast + 4}, \;\;\;
\mu = \frac{i^\ast}{3 i^\ast + 4} \;\;\;\;\;\;\; ({\rm regime \; II})
\end{equation}
\begin{equation}
\label{exp2}
\gamma = \frac{\chi + i^\ast}{i^\ast + 5}, \;\;\;
\mu = \frac{i^\ast+1 }{i^\ast + 5} \;\;\;\;\;\;\; ({\rm regime \; III})
\end{equation}
\begin{equation}
\label{exp3}
\gamma = \frac{\chi + i^\ast}{2 i^\ast + 6}, \;\;\;
\mu = 0  \;\;\;\;\;\;\; 
({\rm regime \; IV}).
\end{equation}

Consider first regime II. Consistency requires the inequalities
$L_c \sim \alpha^\mu \Gamma^\gamma \ll L_\times \sim 
\alpha^{-1/(2 i^\ast - 1)}$ and 
$\alpha^\mu \Gamma^\gamma \ll L_F \sim (\alpha \Gamma)^{1/3} $
to be satisfied simultaneously, which implies 
\begin{equation}
\label{bounds1}
\Gamma^{-(1 - 3 \gamma)/(1 - 3 \mu)} \ll \alpha \ll
\Gamma^{-\gamma/(\mu + 1/(2 i^\ast - 1))}.
\end{equation}
Since $\alpha \ll 1$ and $\Gamma \gg 1$, these inequalities can be 
satisfied only if the exponent of $\Gamma$ on the left hand side 
is smaller than the one on the right hand side. Inserting the 
expressions (\ref{exp1}) this is found to lead to the condition
\begin{equation}
\label{genbound}
\chi < \frac{2}{i^\ast + 1}.
\end{equation}
The analysis of regime III, based on the requirement 
$L_\times \ll  L_c \ll L_F$, yields the same condition
(\ref{genbound}), while in regime IV the 
relevant inequalities $l_{\rm ES} \ll L_c \ll L_F$ translate into
\begin{equation}
\label{weakbound}
\chi < \frac{3 - i^\ast}{2}.
\end{equation}

Using the expression (\ref{chiconv}) for $\chi$, 
we see that the fluctuation-dominated regimes II and III become
inconsistent
at $i^\ast = i^\ast_{c1}= (1 + \sqrt{17})/2 \approx 2.56$. For larger
$i^\ast$ regime III survives in its mean field version, while regime
II disappears. In 
regime IV the transition from fluctuation-dominated to mean-field
behavior occurs at $i^\ast = i^\ast_{c2} = 2$; 
in this regime the critical island size is of the order
of the first layer island distance, $\gamma = \chi/2$,
and hence the problem really
reduces to that of first layer nucleation. 

The general form of the conditions (\ref{genbound},\ref{weakbound}) 
illustrates that the upper critical nucleus size, at which the mean-field nucleation mode sets it, is intimately
linked to the expression for the first layer island
density exponent. 
Scenarios different from the standard case (\ref{chiconv}) can
arise if, due to e.g. metastable clusters \cite{krug00} or 
surface reconstruction \cite{michely96}, 
the critical nucleus on the substrate
is smaller or larger than that on the island. 
Another example of nonstandard behavior is
kinetically limited first layer nucleation, where
Eq.(\ref{chiconv}) is replaced by \cite{kandel}
$
\chi = 2 i^\ast/(i^\ast + 3)
$
and the upper critical nucleus sizes shift to $i^\ast_{c1} = 
\sqrt{3} \approx 1.732$ and $i^\ast_{c2} = \sqrt{13}-2 \approx
1.6055$. In addition the growth law (\ref{growth}) plays an
important role, as will be discussed below.

\subsection{Nonstationary nucleation}
\label{Nonstat}

For very strong step edge barriers the residence time $\tau$ becomes the
largest time scale in the problem, and may effectively be set to infinity
\cite{krug00,heinrichs00}. In this situation all atoms deposited on the
island remain there, and a stationary state where deposition and loss of
adatoms compensate each other is never reached. A distinction between
fluctuation-dominated and mean-field 
nucleation scenarios is nevertheless
possible \cite{heinrichs00}: In the first case  nucleation occurs as soon
as the $i^\ast + 1$-th atom arrives on the island, while in the second
case many more atoms have to be deposited. 

\begin{figure}[htb]
\centering
\vspace*{70mm}
\includegraphics{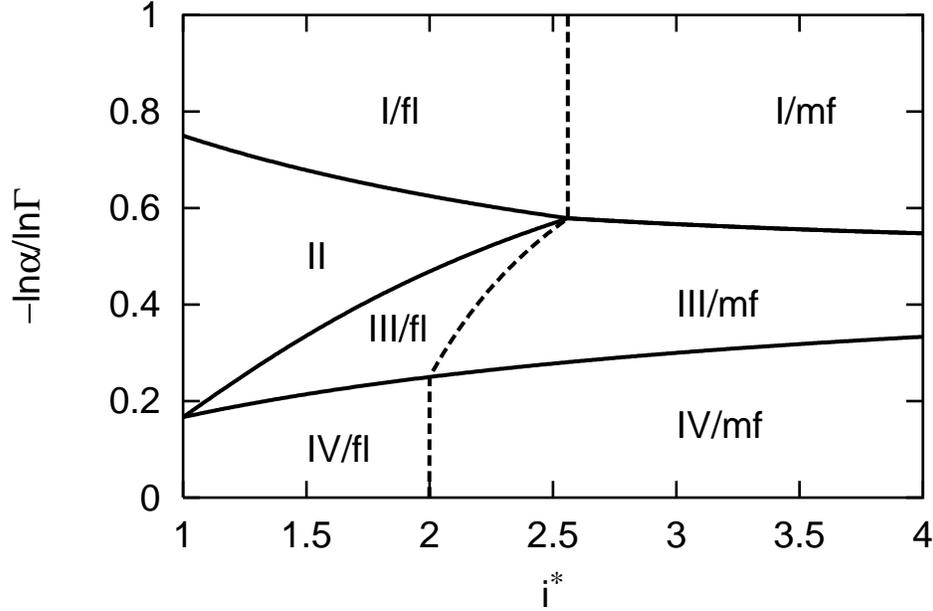}
\vspace*{70mm}
\caption{
Scaling regimes for second layer nucleation in two dimensions.
Each regime is defined by relations of the form
$\Gamma^{-\delta_k} \ll \alpha \ll \Gamma^{-\delta_{k'}}$, 
and the full lines show how the boundaries
$- \ln \alpha/\ln \Gamma = \delta_k$ 
vary with the critical nucleus size $i^\ast$. 
The dashed lines indicate transitions between fluctuation-dominated
(fl)  and mean-field like (mf) subregimes. 
The explicit expressions for 
the boundaries are 
$\delta_1 = 1 - 3 \chi/4$ (I/fl $\to$ II) ;
$\delta_2 = (\chi + i^\ast)(2 i^\ast - 1)/(2 i^\ast 
(i^\ast + 1) + 4)$ (II $\to$ III/fl);
$\delta_3 = (3 \chi + 2 i^\ast - 5)/(2 i^\ast -2)$ (III/fl $\to$ III/mf);
$\delta_4 = (5 + i^\ast - 2 \chi)/(6 + 2 i^\ast)$ (I/mf $\to$ III/mf);
$\delta_5 = \chi/2$ (III $\to$ IV).
For the island density exponent the conventional expression
$\chi = i^\ast/(i^\ast + 2)$ was used.  
}
\label{reg2d}
\end{figure}

Using the growth law (\ref{growth}) it is easy to show that
the time required to deposit a few 
(i.e., $i^\ast + 1$) atoms on the island is
\begin{equation}
\label{deptime}
\tau_{\rm dep} \sim N^{1/2}/F.
\end{equation}
At this point the island size is of the order $N^{-1/4}$,
corresponding to exponents
\begin{equation}
\label{exp4}
\gamma = \chi/4, \;\;\;
\mu = 0,  \;\;\;\;\;\;\; 
({\rm regime \; I/fl})
\end{equation}
for the fluctuation-dominated part of regime I. To see whether
nucleation is actually fluctuation dominated, the time scale (\ref{deptime})
is to be compared to the encounter time (\ref{tauenc}), evaluated
at the island size $N^{-1/4}$; fluctuation-dominated nucleation
occurs if $\tau_{\rm enc} \ll \tau_{\rm dep}$. Once more this yields
the condition (\ref{genbound}), which can therefore be regarded
as a universal criterion for fluctuation-dominated second layer 
nucleation. For $\chi > 2/(i^\ast + 1)$ mean field theory can be used
to compute the nucleation rate, and one arrives at the scaling exponents
\cite{heinrichs00}
\begin{equation}
\label{exp5}
\gamma = \frac{\chi(i^\ast +2) -1}{2 i^\ast + 6}, \;\;\;
\mu = 0  \;\;\;\;\;\;\; 
({\rm regime \; I/mf}).
\end{equation}
A graphical representation
of the sequence of 
scaling regimes as a function of $i^\ast$ is provided in 
Figure \ref{reg2d}.

While in the stationary nucleation regimes of Section \ref{Stat}
the criterion (\ref{genbound}) appears in a rather indirect way,
in the nonstationary case it has a straightforward interpretation
in terms of a comparison of the time scales $\tau_{\rm dep}$
and $\tau_{\rm enc}$. This is easily generalized to first layer
island growth laws of the form
\begin{equation}
\label{growth2}
L(t) \sim N^{-1/2} \theta^\beta
\end{equation}
which lead to
\begin{equation}
\label{nonbound}
\chi < \frac{2 \beta + 1}{i^\ast + 1}
\end{equation}
instead of (\ref{genbound}). 

\section{Dimensionality dependence}
\label{Dimension}

In statistical physics fluctuation effects are typically more
prominent in low spatial dimensionalities. 
It is therefore of some interest to 
repeat the above considerations for growth of $d$-dimensional islands
on a $d$-dimensional substrate. The residence time is then still
given by an expression of the form (\ref{tau}), while the interarrival
time becomes 
\begin{equation}
\label{deltatd}
\Delta t \sim \frac{1}{F L^d}.
\end{equation}
To estimate the encounter time, it is useful to visualize the 
trajectories of the $i^\ast + 1$ diffusing adatoms in a configuration
space of dimensionality $d(i^\ast + 1)$, where the motion is restricted
to a region of linear extent $L$. Nucleation occurs on a $d$-dimensional
submanifold, the codimension of which is $d i^\ast$. We are thus dealing
with a first passage problem in $d i^\ast$ dimensions, where the encounter
time plays the role of the mean first passage time. This implies that 
\begin{equation}
\label{tencd}
\tau_{\rm enc} \sim \left\{ \begin{array}{l@{\quad:\quad}l}
L^2/\nu & d i^\ast < 2 \\ 
L^{d i^\ast}/ \nu  & d i^\ast \geq 2. \end{array} \right.
\end{equation}
Since the residence time is always at least equal to the traversal time
$L^2/\nu$, it follows that encounter-limited nucleation in the sense
of $\tau_{\rm enc} \gg \tau$ is possible only for dimensionalities
$d >  2/i^\ast$. This recovers the upper critical dimension
$d_c = 2/i^\ast$ for diffusion-limited reactions of $i^\ast + 1$
particles \cite{kallabis98,krapivsky94}.

\subsection{Island density exponent in $d$ dimensions}
\label{First}

As a warm-up excercise we derive here a formula for the exponent
$\chi$ of the first layer island density valid for general $d$.
The appropriate expression for the nucleation rate is (\ref{omega3}),
which is now to be interpreted as the rate of nucleation in a region
of size $L$. The distance $l_D \sim N^{-1/d}$ between first
layer nuclei is determined by the condition that there should be 
one nucleation event per area $l_D^d$ during the 
monolayer deposition time $1/F$, i.e. 
\begin{equation}
\label{lD}
\omega(l_D)/F = {\cal{O}}(1).
\end{equation}
For $d i^\ast \geq 2$ this yields
\begin{equation}
\label{chi1}
\chi = \frac{d i^\ast}{d + 2 + 2 i^\ast} \;\;\;\;\;\;\;
(d i^\ast \geq 2),
\end{equation}
which agrees with (\ref{chiconv}) for $d = 2$ and with the
expression $\chi = i^\ast/(3 + 2 i^\ast)$ derived by
Kallabis, Krapivsky and Wolf for $d = 1$, $i^\ast \geq 2$
\cite{kallabis98}. For $d i^\ast < 2$ we obtain instead
\begin{equation}
\label{chi2}
\chi = \frac{d i^\ast}{(d + 2) i^\ast + d} \;\;\;\;\;\;\;
(d i^\ast < 2).
\end{equation}
This case is physically realized only for $d = i^\ast = 1$,
where (\ref{chi2}) reduces to the well-known value
$\chi = 1/4$ \cite{kallabis98,pimp92}.

These results are easily generalized to the growth of islands of
(fractal or integer) dimensionality $d'$ on a
$d$-dimensional substrate. In that case the time of 
interest is not the monolayer growth time $1/F$, but rather
the time required to cover a fraction of the substrate
which is of order unity.
While the true coverage is related to island density and island size
through $\theta \sim N L^{d'}$, the covered fraction 
of the $d$-dimensional substrate is $\theta_d \sim N L^d \sim 
N^{1 - d/d'} \theta^{d/d'}$.  Setting the inverse nucleation rate
equal to the time in which $\theta_d = {\cal{O}}(1)$ is reached then
yields the condition   
\begin{equation}
\label{lDfrac}
\omega(l_D)/F \sim l_D^{d - d'},
\end{equation}
which for $d i^\ast \geq 2$ implies the island density exponent 
\begin{equation}
\label{chigen}
\chi = \frac{d i^\ast}{2(i^\ast + 1) + d'} \;\;\;\;\;\;\;
(d i^\ast \geq 2).
\end{equation}
For $d=1$, 2 this agrees with the general 
expression given in \cite{kallabis98}. 

\subsection{Second layer nucleation in $d$ dimensions}

For compact $d$-dimensional first layer islands the growth law (\ref{growth})
generalizes to 
\begin{equation}
\label{growthd}
L(t) \sim (\theta/N)^{1/d},
\end{equation}
and the critical island size for a nucleation rate of the form
(\ref{Omega}) becomes 
\begin{equation}
\label{Lcd}
L_c \sim (\Omega N)^{-1/(d+k)}.
\end{equation}
Using the expressions (\ref{omega1},\ref{omega2},\ref{omega3}) for the
nucleation rates together with (\ref{deltatd},\ref{tencd}) and 
(\ref{growthd}) it is then
straightforward to repeat the considerations of Section
\ref{Consistent}. Rather than developing in detail
the general case, we focus here on the transition between
fluctuation-dominated and mean field nucleation regimes. 

The
analysis is most transparent in the nonstationary case. One finds
that the deposition time is given by (\ref{deptime}) for all $d$,
and the critical island size for second layer nucleation is
\begin{equation}
\label{regIfl}
L_c \sim N^{-1/2d} \sim \sqrt{l_D} \sim  \Gamma^{\chi/2d}
 \;\;\;\;\;\;\; ({\rm regime \; I/fl}).
\end{equation}
Using this to evaluate the encounter time (\ref{tauenc}), it turns out that
nucleation is always fluctuation-dominated for $d i^\ast < 2$, while
for $d i^\ast \geq 2$ the condition for the fluctuation-dominated regime
is again given by the universal relation (\ref{genbound}). The same relation
also governs the disappearance of the fluctuation-dominated regime II
for general $d$. 

Setting the island density
exponent (\ref{chi1}) equal to $2/(i^\ast + 1)$
then yields the expression
\begin{equation}
\label{ic1}
i^\ast_{c1} = \frac{1}{2d}( 4 - d +  \sqrt{8 d^2 + (d + 4)^2} )
\end{equation}
for the upper critical nucleus size at which the fluctuation-dominated regimes
I/fl, II and III/fl disappear. This is a decreasing function of $d$ which
approaches $i^\ast_{c1} = 1$ for $d \to \infty$. As expected, the 
fluctuation-dominated regime becomes smaller in higher dimensions. 

To analyse the transition to mean field nucleation in regime IV we use
the fact that the critical island size for second layer
nucleation is of the order of the first layer island spacing,
i.e. $\gamma = \chi/d$ and $L_c \sim l_D$. The condition for 
fluctuation-dominated nucleation therefore simply reads
$\tau/\Delta t \sim F l_D^{2 + d}/\nu \ll 1$, or $\chi < d/(2+d)$.
This clearly always holds for the expression (\ref{chi2}), while
for (\ref{chi1}) it is true provided $i^\ast < i^\ast_{c2}$ with
\begin{equation}
\label{ic2}
i^\ast_{c2} = 1 + 2/d.
\end{equation}
Again $i^\ast_{c2} \to 1$ for $d \to \infty$, and
$i^\ast_{c2} < i^\ast_{c1}$ for all $d$.

\begin{figure}[hb]
\centering
\vspace*{70mm}
\includegraphics{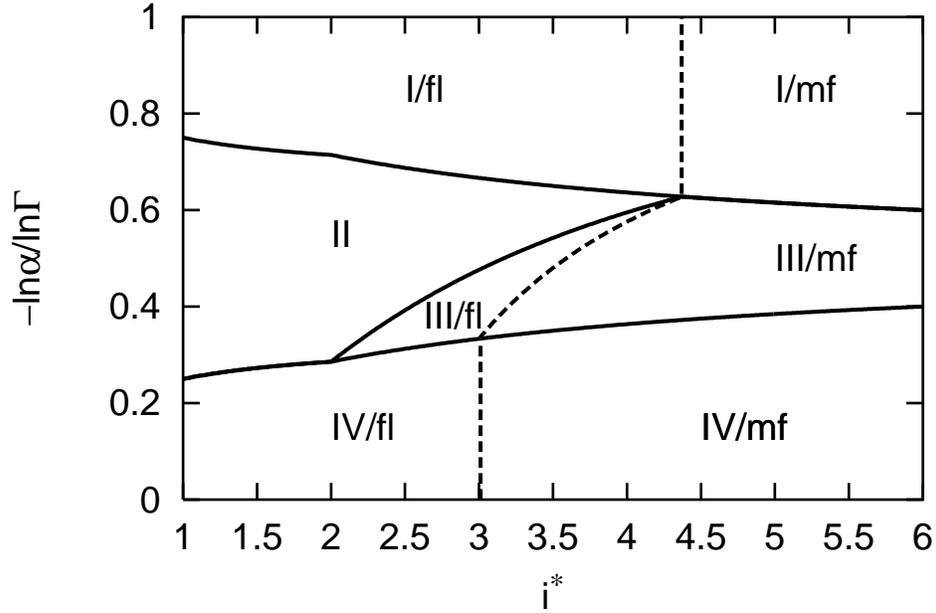}
\vspace*{70mm}
\caption{
Scaling regimes for second layer nucleation in one dimension.
The explicit expressions for 
the boundaries are 
$\delta_1 = 1 - \chi$ (I/fl $\to$ II and I/mf $\to$ III/mf);
$\delta_2 = (\chi + i^\ast)(i^\ast - 1)/(i^\ast 
(i^\ast + 1) + 2)$ (II $\to$ III/fl);
$\delta_3 = 1 - 2(1 - \chi)/(i^\ast -1)$ (III/fl $\to$ III/mf);
$\delta_4 = \chi$ (III $\to$ IV). For the island density exponent
$\chi$ the one-dimensional expressions derived in Section \ref{First}
were used.   
}
\label{reg1d}
\end{figure}

The expressions for the scaling exponents $\gamma$ and $\mu$ in 
the remaining regimes read
\begin{equation}
\label{regId}
\gamma = \frac{\chi(i^\ast + 2) - 1}{d(i^\ast + 3)}, \;\;\;
\mu = 0 \;\;\;\;\;\;\; ({\rm regime \; I/mf})
\end{equation}
\begin{equation}
\label{regIId}
\gamma = \frac{\chi + i^\ast}{(d+1)i^\ast + 2d}, \;\;\;
\mu = \frac{i^\ast}{(d+1)i^\ast + 2d} \;\;\;\;\;\;\; ({\rm regime \; II})
\end{equation}
\begin{equation}
\label{regIIId}
\gamma = \frac{\chi + i^\ast}{i^\ast + 2d+1}, \;\;\;
\mu = \frac{i^\ast+1}{i^\ast + 2d+1}  \;\;\;\;\;\;\; 
({\rm regime \; III}).
\end{equation}
The scaling regimes for the one-dimensional case are shown in Figure
\ref{reg1d}. The upper critical nucleus sizes are 
$i^\ast_{c1} = (3 + \sqrt{33})/2 \approx 4.372...$ and $i^\ast_{c2} = 3$.

\section{Nucleation on the top terrace of a mound}

In multilayer growth, the suppression of interlayer transport
leads to the formation of pyramidal mounds \cite{politi00}. 
For strong step edge barriers, in the sense that $l_{\rm ES}$ exceeds
the distance $l_D$ between first layer islands, the mound separation
is set by $l_D$ and remains constant during growth
\cite{krug97,kalff99}. In this growth
regime the mounds are wedding-cake-like stacks of islands upon islands,
with a characteristic up-down-asymmetry: While the valleys between
mounds are deep crevices, at the hilltops flat terraces of a
characteristic size $L_{\rm top} < l_D$ are found \cite{politi97,kalff99}.
$L_{\rm top}$ is determined by the nucleation rate on the top
terrace through the requirement that on average one nucleation
event should occur during the growth of a monolayer, so that 
\cite{krug00,politi97}
\begin{equation}
\label{Ltop}
\omega(L_{\rm top})/F = {\cal O}(1).
\end{equation} 
For a nucleation rate of the general form (\ref{Omega}) this implies
\begin{equation}
\label{Ltopscal}
L_{\rm top} \sim \Omega^{1/k}.
\end{equation}
A more precise calculation the factor of proportionality can be found
in \cite{krug00}.

As in the case of second layer nucleation, application of the expressions
(\ref{omega1}, \ref{omega2}, \ref{omega3}) for the nucleation rate
yields estimates for the top terrace size of the form
\begin{equation}
\label{Ltopgen}
L_{\rm top} \sim \Gamma^{\gamma'} \alpha^{\mu'}
\end{equation}
where the scaling exponents $\gamma'$ and $\mu'$ depend on 
$i^\ast$ and on the scaling regime in question. Inspection shows
that the equivalent of the nonstationary regime I discussed above
in Section \ref{Nonstat} is trivial in the case of mound growth:
When $\alpha \ll \Gamma^{-1}$, the top terrace size becomes comparable
to the lattice constant, $L_{\rm top} = {\cal O}(1)$, corresponding
to pure statistical Poisson growth \cite{krug97}. 

The fluctuation-dominated regime II with 
scaling exponents  
\begin{equation}
\label{moundsII}
\gamma' = \mu' = \frac{i^\ast}{3 i^\ast + 2}, 
\;\;\;\;\;\;\; ({\rm regime \; II, \;\; mounds})
\end{equation}
covers the parameter range $\Gamma^{-1} \ll \alpha \ll
\Gamma^{-\delta_1}$, where $\delta_1 = 
i^\ast(2i^\ast - 1)/(2i^\ast(i^\ast + 1) + 2) < 1$ for all $i^\ast$.
In marked contrast to second layer nucleation, here regime II 
exists for all $i^\ast$. 

\begin{figure}[hb]
\centering
\vspace*{70mm}
\includegraphics{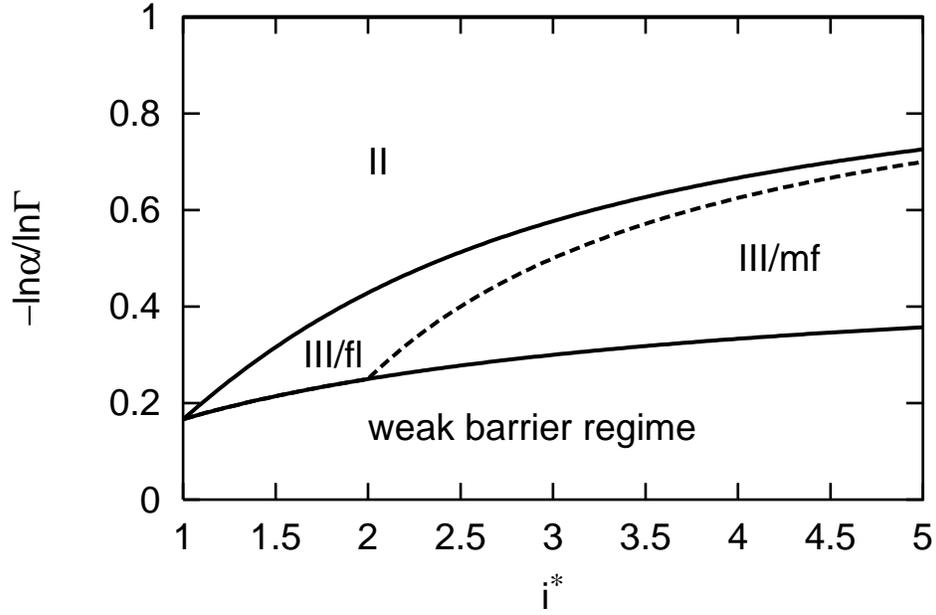}
\vspace*{70mm}
\caption{
Scaling regimes for nucleation on top of two-dimensional mounds.
The explicit expressions for 
the boundaries are 
$\delta_1 = (2i^\ast - 1)i^\ast/(2i^\ast(i^\ast + 1) + 2)$ (II $\to$ III/fl);
$\delta_2 = 1 - 3/2i^\ast$ (III/fl $\to$ III/mf);
$\delta_3 = \chi/2$ (III $\to$ IV).  
}
\label{moundreg}
\end{figure}

For $\Gamma^{-\delta_1} \ll \alpha \ll
\Gamma^{-\chi/2}$ nucleation is governed by the exponents of regime III,
\begin{equation}
\label{moundsIII}
\gamma' = \frac{i^\ast}{i^\ast + 3}, \;\;\;
\mu' = \frac{i^\ast + 1}{i^\ast + 3} 
\;\;\;\;\;\;\; ({\rm regime \; III, \;\; mounds}).
\end{equation}
This regime is fluctuation-dominated for $i^\ast < 2$, while
for $i^\ast \geq 2$ it splits up into fluctuation-dominated and
mean field-like subregimes with identical scaling exponents. 
Both subregimes exist for all $i^\ast \geq 2$, though the
fluctuation-dominated part becomes very small for large $i^\ast$.
For $\alpha > \Gamma^{-\chi/2}$ one enters the regime of weak step 
edge barriers, where $l_{\rm ES} < l_D$ and the 
shape and evolution 
of the mound morphology are qualitatively different 
from the wedding-cake regime discussed here
\cite{politi96,politi97,politi00}.
A pictorial representation of the scaling
regimes is given in Figure \ref{moundreg}.

\section{Summary} 

The goal of this paper has been to extend the analysis of \cite{krug00}
to arbitrary critical nucleus sizes, as well as to clarify the relationship
between the recent stochastic approaches to second layer nucleation 
\cite{rottler99,krug00,heinrichs00} and the earlier mean field theory
of Ref.\cite{tersoff94}. The essential new ingredient that enters the problem
for $i^\ast > 1$ is the encounter time $\tau_{\rm enc}$, which may
become larger than the residence time $\tau$. The mean field approach is
valid, at least as far as the values of scaling exponents are concerned,
when $\tau \ll \tau_ {\rm enc}$, but it fails when $\tau \geq \tau_{\rm enc}$.
Apart from recovering the results of \cite{heinrichs00}, the analysis
has revealed a remarkably simple and universal criterion for fluctuation-dominated second layer nucleation, Eq.(\ref{genbound}),
which may be useful in situations where nucleation is governed
by different mechanisms in the first and second layer. Furthermore I have
verified the expectation that fluctuation-dominated nucleation should
be more prominent in low dimensionalities, and I have shown that 
no upper critical nucleus size (beyond which fluctuations can be neglected)
exists for nucleation on the top terrace of a mound.

\section*{Acknowledgements} I am grateful to S. Heinrichs, P. Maass
and T. Michely for useful discussions. This work was supported by
DFG within SFB237.

 \end{document}